# Double Whammy –
# How ICT Projects are Fooled by Randomness and Screwed by Political Intent

By

Alexander Budzier and Bent Flyvbjerg

BT Centre for Major Programme Management

Saïd Business School

University of Oxford

## Abstract

The Iron Triangle formulates the holy trinity of objectives of project management – cost, schedule, and benefits. As our previous research has shown, ICT projects deviate from their initial cost estimate by more than 10% in 8 out of 10 cases. Academic research has argued that Optimism Bias and Black Swan Blindness cause forecasts to fall short of actual costs. Firstly, optimism bias has been linked to effects of deception and delusion, which is caused by taking the inside-view and ignoring distributional information when making decisions. Secondly, we argued before that Black Swan Blindness makes decision-makers ignore outlying events even if decisions and judgements are based on the outside view. Using a sample of 1,471 ICT projects with a total value of USD 241 billion – we answer the question: Can we show the different effects of Normal Performance, Delusion, and Deception?

We calculated the cumulative distribution function (CDF) of $(actual - forecast)/forecast$. Our results show that the CDF changes at two tipping points – the first one transforms an exponential function into a Gaussian bell curve. The second tipping point transforms the bell curve into a power law distribution with the power of 2.

We argue that these results show that project performance up to the first tipping point is politically motivated and project performance above the second tipping point indicates that project managers and decision-makers are fooled by random outliers, because they are blind to thick tails. We then show that Black Swan ICT projects are a significant source of uncertainty to an organisation and that management needs to be aware of.

Finally, we draw implications about the underlying generative processes that lead to power law behaviour, which might help to further understand the pitfalls and shortcomings of cost and cost risk management in ICT projects.

“Marx referred to technology as 'frozen labour' - work and its values embedded and inscribed in transportable form... where they [technology] are used to make decisions, or to represent decision-making processes, such technologies also act to embed those decisions. That is, the arguments, decisions, uncertainties, and processural nature of decision-making are hidden away inside a piece of technology or in a complex representation. Thus values, opinions, and rhetoric are frozen into codes, electronic thresholds and computer applications. Extending Marx, then, we can say that in many ways, software is frozen organisational discourse.” (Bowker & Star 1994:187)

“For the past 40 years, for example, we’ve tortured ourselves over our inability to finish a software project on time and on budget. But as I hinted earlier, this never should have been the supreme goal. The more important goal is transformation, creating software that changes the world or that transforms a company or how it does business. We’ve been rather successful at transformation, often while operating outside our control envelope. Software development is and always will be somewhat experimental. The actual software construction isn’t necessarily experimental, but its conception is. And this is where our focus ought to be. It’s where our focus always ought to have been.” (DeMarco 1995:95)

# A Tale of Two Projects

In May 1996 the decision-makers at the Benefits Agency of the UK Department of Social Security and the executives at Post Office Counters Ltd. jointly made a decision. They awarded a one billion GBP, seven year contract to Pathway, a subsidiary of Fujitsu. Their decision would finally teleport the paper-based benefits payment process into the 21st century by January 1999. This decision was a major transformation, it would change how 20,000 post offices would work, and it would change how 17 million benefits recipients would receive their money, a number that sums up to about 760 million payments every year. However, the decision quickly turned sour - in July the first issues were discovered two month before the system development was planned to start, by the end of the year the project was re-baselined to allow increase the time for system development from seven to twelve months. In the end the project was abandoned in May 1999, software development was not finished and the most current forecast projected the project to deliver three years behind scheduled go-live, 30% budget overrun and a total spend of 1 billion pounds (National Audit Office 2008).

The private sector does not fare better in comparison. In 2003 the Levi Strauss Company celebrated its 150 birthday. Apart from the company's birthday the management was facing two highly critical issues. First, international expansion to over 110 countries had lead to a highly fragmented and overlapping IT architecture. The company operated a Baan system in Europe and ran a mix of bespoke mainframe solutions for Asia, Canada, and the US. Secondly, the 2002 Sarbanes-Oxley Act put pressure on auditors and financial processes. A new auditor revealed flaws in accounting procedures, which almost forced Levi Strauss to restate their accounts. Presented with a burning platform the top-management decided to revamp the complete architecture. The project aimed to implement SAP with the help of Deloitte consulting. Analysts estimated the project cost to be USD 5 million without consulting fees. Project complexity increased quickly: while the project was being first implemented in Asia, Levi-Strauss' key client Walmart required that the existing - and soon to be replaced - systems interface with their supply chain management system. In 2008 the system was finally rolled out to the U.S. market. Project management anticipated risks during switch-over. The company pre-shipped as many Q2 orders in Q1 as possible. However, when the actual go-live happened the three

US distribution centres went offline for a full week. As a result Levi Strauss reported a USD 192.5 million hit to the bottom line for Q2. David Bergen, who had been Levi's CIO for 8 years, resigned from his post. SAP still remains the key vendor to Levi Strauss (ZDNet, SAP, The Register, ITPro).

The Benefits Cards Payments project and Levi Strauss' ERP debacle clearly raise the question: Are these two tales archetypical for the performance of ICT projects or are they one-of-a-kind bad apples?

## How Risky are ICT Projects Really?

Several attempts have been made to turn the anecdotal evidence of cost and schedule overruns and benefits shortfalls in large-scale ICT projects into surveys that measure risk more systematically.

When the Standish Group first published the now infamous Chaos Report in 1999 their findings reflected the perceived high risk of ICT projects. Similarly the Function Point Estimation work done by Jones (Jones 1998) reported a plummeting success rate of ICT projects with increasing size. Both studies saw virtually no success for projects larger than 6 million dollars.

However, academia contested these findings on the grounds that sampling was skewed towards failure, data collection was opaque, and the categorisation of success, challenged, failed projects was methodologically biased (Eveleens & Verhoef 2010; Keil 1995, Keil & Mann 1997, Keil, Mann & Rai 2000b; Jorgensen & Molokken-Osvold 2006, Molokken-Osvold & Jorgensen 2003; Sauer et al. 2007; Zhang, Keil, Rai & Mann 2003).

But then how risky are ICT projects really? Three approaches can be found in the existing literature (1) studies of the effectiveness of software cost/scope/schedule estimation techniques; (2) surveys of software professionals, and (3) analyses of ICT project portfolios. Table 2 gives an overview of these studies, while the following discusses key contributions.

The first approach to measure the risks associated with ICT projects is anchored in the field of software development metrics. Jones' (1998, 2003) work on function points re-presents the longest academic record of these types of analyses. While the dataset itself only measures schedule risk (average of -8%) it gives an indication of the failure and success rate. Jones reports a success rate of 14% (Jones 2008) and 18% (Jones 1998) and a cancellation rate of 13%, 24%, 29% for 2000, 2007, and 2008 respectively. While Jones' work was fundamental for establishing metrics of software quality and productivity the data do not reflect overall project performances. Firstly, because patterns of productivity, quality and scope creep de-stabilise above a scope of 10,000 function points (approx. 30 months of project duration). Secondly, the methods cover soft-ware development explicitly, incidental efforts such as overtime, documentation, project management, and change to business processes constitute up to 75% and estimates for the same project differ for up to 100% depending on which of the 10 available function point methods is used as estimator (Jones 2008). Moløkken & Jørgensen (2003; 2006) conducted a meta-analysis of 10 software estimation surveys published between 1984 and 2002. These 10 studies reported cost overruns between 33-34% and 22% schedule overrun. Budget and schedule overruns occurred in 6-7 out of 10 projects, under-runs in 1 out of 10.

A second approach to measure the ICT project risk is surveys of project professionals. Keil et al. (Keil, Tan, Wei, Saarinen, Tuunainen & Wassenaar 2000a; Keil & Flatto 1999; Keil 1995; Montealegre & Keil 2000; Keil & Mann 1997a; Zhang, Keil, Rai & Mann 2003) surveyed IS auditors and asked how many of the last projects, they were involved with, escalated and by how much. The survey found that 30-40% of all projects escalated on average by 21 months, median 15 months. Non-escalated projects delivered on average 18% over budget, and 22% behind schedule; whilst escalated projects come in 156% over budget and 133% behind schedule. Sauer et al. (Sauer et al. 2007; Sauer & Cuthbertson 2004) surveyed 412 project managers about their current or most recent projects. The survey reported that a reasonable expectation of risk is 7% cost, schedule overrun and scope shortfall in 2 out of 3 projects. Studying ICT project performance with survey-based methods introduces the considerable memory bias of which two aspects are particularly problematic (1) consistency/desirability bias and (2) recency effects. Firstly, most of these studies have been framed in the context of the inflated Standish Figures (Anon 1999), which were mostly perceived as unjustly distorting the real performance of ICT project managers (Glass 2006). Thus motivating the subjects to lean towards a more positive view of project performance. Secondly, the human memory biases, over-emphasises recent experiences and critical events (Chell 2004) over of the average performance, a bias that would motivate respondents to underreport project performance.

A third approach to quantify ICT project risk has been pursued by Verhoef and his research group who first used the Jones (Jones 1998) data to model risk quantitatively (Verhoef 2002). In two studies (Eveleens & Verhoef 2009; Kulk et al. 2009), the group analysed the ICT project portfolio of two organisations that consisted of 1,172 projects, the project size was on average 581,000 EUR. The first study reported 286% average cost overruns for organisation 1 (n=867 projects), median cost under-run of 38%. For the second organisation (n=140 projects) cost overruns were 16% on average, the median cost under-run was 2%. The functionality delivered was only measured in organisation 2 where projects over-delivered by 10% on average, median 0% (Eveleens & Verhoef 2009). These two studies reported largely skewed cost overrun data with half the portfolio coming in under budget and a much larger average than median. The data indicates that the majority of projects stays within the budget but the considerable right-tail skew typically results from right-tail outliers. The second study (Kulk et al. 2009) looked at one additional organisation with 165 projects in total with an average project budget of 2.2 million EUR. The average cost overrun is about 0%, they argued that a good estimate should forecast a budget that is [-5%, 2.5%] around the actual cost. Furthermore the two studies found that the five organisations showed unique pattern of accuracy, biases, and patterns of convergence of forecasts to actual figures (Eveleens & Verhoef 2009; Kulk et al. 2009). These studies addressed the previous issues of selective reporting and respondent biases effectively. However, as the authors pointed out sampling from a limited number of firms and the resulting very different patters for the each of the five organisation make cross comparison impossible either due to organisational, industry or geographic differences. Moreover the studies did not check for confounding effects, such as project size or duration. Table 1 gives an overview of previous studies of ICT project risk.

However, when looking at the demand or buying side of the ICT project a whole different picture emerges. In a non-academic survey CFOs were asked about their ICT project performance. The survey reports that only 35% of all finance ICT projects delivered on budget, with 27% running <15% over budget; 16% of the projects between 15-50% over budget and another 16% with more than 50% budget overrun. Only 7% of the projects delivered on time, 25% of the project showed slight delays, 68% of the projects showed major delays (Friedman & van Decker 2010).

In sum the academic studies have not found support for alarmist statistics. Even though the scale is more than an order of magnitude smaller deviations from the initially planned forecasts have been observed. Which raises the question what are the root causes of the risk?

| Author | Comments | Year | Sample drawn from # organisations/ participants | Method | Effective Sample Size (# of used projects) | Average Actual Size of the project | Median Actual Size of the project | Success | | | Cost/Effort | | | Schedule | | | Functionality/Benefits | | |
|---|---|---|---|---|---|---|---|---|---|---|---|---|---|---|---|---|---|---|---|
| | | | | | | | | Successful | Challenged | Failure | Average | Median | Standard deviation | Average | Median | Standard deviation | Average | Median | Standard deviation |
| Addison & Vallabh | no data | 2002 | 36 | no data collected only risk factor | 36 | | | | | | | | | | | | | | |
| Augustine (cited in Eveleens, 2011; Bergeron & | n/a | 1979 | | | 100 | | | | | | | | | 33% | | | | | |
| Bergeron & St-Arnaud | | 1992 | 67 | questionnaire was used to retros | 89 | 1251 PD | 657 PD | | | | 33% | | | | | | | | |
| Conte et al. | no data, but proposed a-f/f formula | 1986 | | | | | | | | | | | | | | | | | |
| Computerweekly et al. (2003) | | 2003 | | | | | | 27% | 73% | 5% | 18% | | | | | | | | |
| Eveleens & Verhoef | no combined data | 2009 | 4 | | 1824 | | | | | | | | | | | | | | |
| Eveleens/Little (Landmark 2002) | | 2009 | 1 | | 121 | | | | | | | -15% | | 105% | 79% | | | | |
| Eveleens (Org X - large MNO) | | 2009 | 1 | | 867 | | | | | | 286% | -38% | | | | | | | |
| Eveleens (Org Y - Fin Serv) | | 2009 | 1 | | 168 | | | | | | 16% | -2% | | | | | 10% | 0% | |
| Eveleens (Org Z - TelCo) | | 2009 | 1 | | 307 | | | | | | | 0% | | | | | | | |
| Glass | only Jorgensen's data | 2005 | | | | | | | | | | | | | | | | | |
| Glass | no data | 2006 | | | | | | | | | | | | | | | | | |
| Heemstra | | 1989 | 598 | | 2659 | | | | 70% | | | | | | | | | | |
| Huber (Sauer & Cuthbertson) | | 2003 | 421 projects, 1500 partici | Survey of IT Project Managers | | | | 16% | 75% | 9% | 18% | | | 23% | | | -7% | | |
| Jenkins et al. | Average=median effort overrun 36% | 1984 | 23 | Project records, interviews | 72 | | 103k USD | | 61% | | 67% | 34% | | 22% | | | | | |
| Jones | success/failure determined by schedule slip | 2007 | | Data Collection through software size assessments | | | | 62% | 14% | 24% | | | | | | | | | |
| Jones | Scope Creep probability 28.4% (10k FP); 3 | 2008 | | Data Collection through software | 8290 (800 since 2001; 1,385 are maintenance) | | | | | 29% | | | | -8% | | | | | |
| Jones | | 2000 | | Data Collection through software | 9,155 (of which 560>10k FP) | | | 69% | 18% | 13% | | | | | | | | | |
| Kampstra & Verhoef | using Jones data | 2009 | | Using Jones (2000) data to predict failure | | | | | | | | | | | | | | | |
| Keil et al. | Perceived overrun | 2000 | 579 (422 escalated + 157 | Survey of auditors | 579 | | | | 38% | | 119% (average) 156% (escalated) and | | | 103% (average), 133% (escalated) and 22% (non-escalated) | | | | | |
| Kulk et al. | Misestimation is (-5%;2.5%) | 2009 | 1 | Portfolio data | 165 | 2.24 mio USD | | | | | -4% (e) | -10% (e) | | | | | | | |
| Lederer | | 1991 | | | Total 1,622 pers | mostly small projects | | | 60% | | | No average given | | | | | | | |
| Little | | 2006 | 1 | Software development metrics | 106 | 329 days | | | | | 80% | 100% | | | | | | | |
| McAulay | Data on Standards Use only | 1987 | 120 | | 280 | | | | | | | | | | | | | | |
| McKeen | | 1983 | 5 | Random draw of application syst | 32 | 35,000 USD | 40,000 USD | | | | 28% | 22% | | 18% | 22% | | | | |
| Moløkken-Østvold et al. | No size effect found; expert vs. combination | 2004 | 11 | Interview of the project manager | 36 | 3124.5 PH | 1175 PH | 11% | 84% | 5% | 41% | 21% | | 25% | 9% | | | | |
| Moores | Use of Software Estimation Techniques | 1992 | 54 | | 115 | | | | | | | | | | | | | | |
| Phan | No data | 1988 | 191 | | 827 | | | | | | 33% | | | | | | | | |
| Sauer et al. | | 2007 | | | 412 | | | | | | 13% | | | 20% | | | -7% | | |
| Standish | Measuring the size of overruns when they occur | 1994 | 365 | | | | | | | | 189% | | | 222% | | | | | |
| Standish | | 1996 | n/a | | | | | | | | 142% | | | 131% | | | | | |
| Standish | | 1998 | n/a | | | | | | | | 69% | | | 79% | | | | | |
| Standish | | 2000 | n/a | | | | | | | | 45% | | | 63% | | | | | |
| Standish | | 2002 | n/a | project profiles, project tracking, individual project surveys, case interviews, general surveys, project post-mortem, and other instruments | 80,000 IT projects over 16 years (5,000 per survey) | The reported resolution and the split by size indicates that average must be 3-5mio USD range | The median must be 1-3m USD | 34% | 51% | 15% | 43% | | | 82% | | | 67% | | |
| Standish | | 2004 | n/a | | | | | 29% | 53% | 18% | 56% | | | 84% | | | 64% | | |
| Standish | | 2006 | n/a | | | | | 35% | 46% | 19% | 47% | | | 72% | | | 68% | | |
| Standish | | 2008 | n/a | | | | | 32% | 44% | 24% | 54% | | | 79% | | | 67% | | |
| Standish | | 2010 | n/a | | | | | 37% | 42% | 21% | 46% | | | 71% | | | 74% | | |
| Topping (cited by Bergeron) | not found | 1985 | | | 22 | | | | | | 40% | 26% | | | | | | | |
| Verhoef | using Jones data | 2002 | | | | | | | | | | | | | | | | | |
| Wydenbach | No data, surveyed use of estimation metho | 1995 | 213 | | 515 | | | | | | | | | | | | | | |

## Black Swan Blindness as a new cause of ICT project risk

The notion of risk in ICT projects is typically rooted in a normative-positivist conception that implies that all risks are manageable (Stahl et al. 2006). The most commonly used definition of ICT risk is McFarlan's (1981:143): “...exposure to such consequences as: failure to obtain all, or even any, of the anticipated benefits; costs of implementation that vastly exceed planned levels; time for implementation that is much greater than expected” McFarlan's definition mirrors the iron triangle of cost-schedule-benefits as the key performance dimensions of projects, the author follows the notion that other commonly used outcome variables such as scope, user satisfaction, should ultimately contribute to the organisation's bottom-line. A second often cited definition of risk is “Exposure=Probability*Loss” (Boehm 1991:147). The causes for risks clearly lie in our incomplete knowledge of the subject matter, thus if a project establishes all possible causes of risks they can be managed away. Fortune & White (2006) used a meta-analysis of more than a decade of research to compile a ‘complete’ list of risk factors. In total the authors distilled 27 critical success factors for ICT projects (Fortune & White 2006).

An alternative school of thought acknowledges more clearly the social construction of ICT project risks. A turning point in software engineering was the discovery that ICT projects are 'Peopleware' (DeMarco & Lister 1999), sometimes also explained within the phrase “ICT systems are social systems” (Crewe & Young 2002:13). The same acknowledgement of ICT systems being social systems is made for instance by Mumford (1995), and earliest by Brooks (1975), who pointed out the detrimental effects of adding programmers to a delayed projects and the impact of overhead and incidental activities on the productivity of programmers. The acknowledgement of the social side of projects puts risk management into the social sciences. Who look for the causes of risks into individual and group processes, e.g. escalation of commitment (Staw & Ross 1989), groupthink (Janis 1971), or principal-agent behaviour (Demski & Feltham, 1978).

A third school of thought is based on the foundational work of Behavioural Economics. Kahneman & Tversky (1977) asked: “What factors limit intuitive judgements and how to correct them?” The authors found that errors of judgement are systematic. Ex-perts and laypeople make similar errors because they neglect distributional data and take an internal approach. Moreover, they found that intuitive predictions are typically non-regressive. Thus expert estimates are usually overconfident in the absence of distributional and conditional data. Lastly, judgement and decision-making from experience is usually limited to the 'normal' condition and heavily influenced by anchoring, e.g., by plans and recent experiences (Kahneman & Tversky 1977). In sum, behavioural economics argues that intuitive judgements are systematically biased towards optimism and overconfidence. However, in a review of 15 studies on the performance of expert estimates vs. model estimates Jørgensen (2004) finds no conclusive evidence for either method. Moreover, he argues, the results indicate that expert estimates in software development are better performing than expert estimates in other professions. Jørgensen explains his findings with the high volatility of ICT projects, which prefers context-rich expert estimates over formulaic models, and the poor performance of formal estimation methods compared to estimation methods in other professions, due to low correlations between schedule, cost, and effort.

Organisational behaviour analysis found that managerial optimism biases are based on the illusion of control (March & Shapira 1987). Illusion of control is described as the "expectancy of a personal success probability inappropriately higher than the objective probability would warrant" (Langer 1975:311). The illusion of control is ubiquitously institutionalised through cybernetic control systems, e.g. accounting, budgeting, KPI performance measurements, Balanced Score Cards (Dermer & Lucas 1986). On an individual managerial level these delusions have not only been explained by the illusion of control but also by the escalation of commitment (Staw 1981, Keil & Mann 1997, Drummond 1999) and the self- and organisational desirability of a course of action (Heaton 2002; Staw 1981; Budescu & Bruderman 1995; Klein & Helweg-Larsen 2002).

Based on their earlier work, Kahneman & Lovallo (1993) re-visit the inside view to unpack the concept of uniqueness and its relation to decision-making. Their conceptual argument was that isolation errors stem from perceived uniqueness: forecasts are over-optimistic because they are anchored on plans and scenarios of success, while singular decisions neglect the option to pool risks and thus tend to be overly timid. Thus, Kahneman & Lovallo (1993) argue that taking the inside view to approach managerial problems is another explanation (on top of the illusion of control) of managerial optimism. Flyvbjerg et al. (2002; 2005; 2007) further argue that not only managerial delusions of success are the bases of optimism bias but also strategic deception (Flyvbjerg et al. 2002). In the empirical work on construction projects the authors found a relatively constant level of cost overruns, which they link conceptually to an interaction effect of deception and delusion, whereby delusion biases are gradually replaced by deceptive biases with increasing organisational and political pressures (Flyvbjerg, Bruzelius & Rothengatter 2006).

This paper offers a new explanation of ICT project risk: "Black Swan Blindness". Recently the risk management literature starts referring to high impact, rare events as Black Swans - a term which has been coined by Taleb (2005; 2007). In the wider context of social studies of risk management Black Swans have been located within the wider field of unknown-unknowns next to Wicked Problems and Post Normal Science (Elahi 2011). According to Elahi's interpretation of unknown-unknowns, Black Swans are characterised by their improbability of happening and the human tendency to post-rationalise their existence.

Organisational behaviour studies show that decision-makers are Black Swan myopic. March & Shapira (1987) found that managerial decision-makers tend to ignore rare, low probability events and overestimate common events. March & Shapira argue based on two case studies that risk is not conceived as probabilistic with the possible impact being more salient than the likelihood of the risk happening. Furthermore they found that attitudes towards risk are stronger determined by managerial factors (e.g., attention, illusion of control, reputation concerns) than systemic, organisational factors. Managers ignore rare events and overestimate common events because of a focus on short-term rewards for which known risks seem controllable. Most interestingly, March & Shapira find that in cases of success managers undertake elaborate strategies to inflate the perceived riskiness and play down the role of chance, implying that the rules of chance do not apply to them or that the rules can be changed (March & Shapira 1987). Their findings also raise a considerable concern to the ability of success case narratives, given that organisational informants are re-rationalising past decisions and report cases self-consistent (Lecky 1945).

Dutton & Webster (1988) offer a different explanation for ignorance of risks: reduced importance. Their study found that high uncertainty, as embodied by Black Swans, reduces the importance of an issue for decision-makers. Both studies align with Elahi (2011) conceptualisation of unknown-unknowns, which tend to be treated with ignorance in organisations because they are uncomfortable or even forbidden organisational knowledge (Stirling 2010). However, Laughunn et al. (1980) observed that decision-makers do pay attention to potentially ruinous risks. 44% of their studied managers switched to a risk-averse behaviour when ruinous risks where introduced into experiments. Following the same argument of potential ruin, Meszaros (1999) takes experimental findings into the wild and studies decision-making with potentially ruinous risks in six chemical firms in Philadelphia. The informants recounted organisational histories of how firms re-assessed risks after the 1984 Bhopal catastrophe. Meszaros constructs a nested survivability heuristic. The heuristic first asks the question whether a rare event is a threat to the survival of the company, followed by the question whether dealing with the risk appeals to upper management, and finally whether an adequate response to the risk is affordable to the company (Meszaros 1999).

In sum, this paper argues that next to social processes that lead to optimism bias in form of deception and delusions, risks are underestimated if the risk distribution has fat tails, thus if the decision at hand turns out to be full of outliers, so called 'Black Swan Events'.

## Hypotheses of three regimes of project performance

The dispute of unacceptable cost overruns vs. acceptable risk could easily be discredited as a debate of experts vs. laymen, of anecdotal vs. systematic evidence — similar to the debate about nuclear safety (Wynne, 1996). In contrast, this paper is going to propose a novel hypothesis: Both accounts of ICT project risk can be supported by data. We propose this first hypothesis behind our reasoning:

- H1: ICT project risk is a heavy tail distribution where outliers, i.e. out-of-control Black Swan project, hide behind good average performance.

This paper further argues that

- H2: Outlying ICT projects follow a different statistical regime than normal projects.
- H2a: ICT projects that come in under budget (left tail outliers) do so not because of performance achievements but because their budgets were cut.
- H2b: ICT projects that come in well over budget (right tail outliers) do so because they spun out-of-control and turned into Black Swans.

Next this paper is going to discuss the data used for the study. Then we test H1 and H2, analysing the three regimes (1) normal Gaussian bell-curve performance, (2) politically influenced budget cuts, and (3) out-of-control cost overruns. Then the paper is sizing the impact of out-of-control projects and finally gives a short overview of causes of Black Swans.

## Project Archaeology

The data for our study was gathered in a guided project archaeology. The questionnaire design followed the methodology used in Flyvbjerg, Holm & Buhl (2002), Flyvbjerg, Skamris, Holm & Buhl (2005), and Flyvbjerg (2007) to collect project estimates for cost, schedule, and benefits along key decision-points in a project. We asked the participating organisations to provide data on the last 20-30 finished projects with as many data points on key decision within the project as possible. Participating organisations have been alerted to take utmost care in checking the figures provided against the original decision documents or the figures stored in their portfolio and project management or audit systems. In the worst case of organisational forgetting only the estimated costs and schedule at the green-light decision and the actual cost and schedule were available. We approached around 200 private sector organisations of which 20 participated, which results in a response rate of 10%. In several cases organisations expressed interest and the data collection was started but yielded no results because no recorded information could be found. This archaeology resulted in a total sample of 142 ICT projects with rich data.

At the same time we sampled from publicly available audit reports from the U.S. Government Accountability Office (GAO) and the U.K. National Audit Office (NAO). We were able to collect 149 large-scale ICT projects from the auditor's reports.

The third data source we used was the United States Federal Budget Data. Each ICT project is required to file an E300 with the White House's Office of Management and Budget. We collected as many of those forms as possible across all 26 agencies and their 134 bureaus within the United States federal government. Our efforts resulted in 13,166 forms, of which 80% automatically scanned and 20% had to be entered by hand. This resulted in a dataset of 5,137 projects. We regard projects as finished that have used at least 90% of their budget when the report was filed, which yielded a total of 2,555 projects. To observe changes in forecasts we had to further narrow the project sample down to the multi-year projects, which resulted in a final sample of 1,180 projects.

Thus in total our sample comprises 1,471 projects, which represents a total value of USD 241 billion (in 2010 USD), it is the largest academic dataset to date. The average project size is USD 122.1m (plan) and USD 167.1m (actual) respectively. The median project size is USD 3.3m (plan) and USD 5.8m (actual). Table 2 details the characteristics of the projects in our sample. Table 2 shows that our data spans not only software development projects but a broad array of project types; the table also shows that among the software implementation projects we find a broad range of ICT systems. Lastly, it shows that our sample substantially covers the decision years 2005-2010.

Table 2 Description of Sample

| **Project Type** | **Frequency** |
|---|---|
| IT Integration | 1% |
| Standard Software | 7% |
| Bespoke Software | 72% |
| IT Infrastructure | 8% |
| IT Architecture | 9% |
| Other | 12% |
| **System Type** | **Frequency** |
| ERP | 28% |
| MIS | 25% |
| Office Systems, e.g. DCM | 9% |
| Disposition, e.g. CBS, SCM | 12% |
| Transaction | 2% |
| Other | 24% |
| **Year of Green-Light Decision** | **Frequency** |
| 2004 and earlier | 5% |
| 2005 | 11% |
| 2006 | 3% |
| 2007 | 12% |
| 2008 | 47% |
| 2009 | 12% |
| 2010 | 9% |

## ICT project risk distribution has fat tails

The first hypothesis postulated that ICT project risk is a fat, heavy tailed distribution. The table 3 lists the moments of the ICT project sample.

Table 3 Risk distribution

| Distribution Moment | Value |
|---|---|
| Average cost risk | 26.74 |
| Standard deviation | 82.23 |
| Skewness | 3.76 |
| Kurtosis | 18.79 |

While the average cost risk seems acceptable the standard deviation is comparatively large, indicating that in a normal distribution of this mean and standard deviation 95% of all observations would fall within (-100%, +188%). However the third and fourth order moment clearly show that this distribution is far from normality.

Skewness is the third central moment and a measure of symmetry. The positive value of 3.76 indicates a right skew in the distribution that is twice as large as the right-tail skew in the construction project sample.

Kurtosis is the fourth order moment and a measure of centrality. Kurtosis of 0 indicates normal distribution. The higher positive kurtosis shows that the distribution is more clustered around the centre of the distribution of ICT cost risks than in the distribution of construction cost risks. Thus the ICT risk distribution shows thinner tails but only until the outlier values, at which the details are much thicker than normality suggests. Thus skewness and fat tails are more present in the distribution of cost risk in ICT than construction, which again is more skewed and has fatter tails than the normal distribution.

We follow Taleb (2007) in our conceptualisation that Black Swans are high impact, rare events. Thus we expect Black Swans to translate into statistical outlier and extreme values. We tested whether outliers might be present. The existence of outliers has been noted before for instance in Kulk et al. (2009). While analysing sizeable ICT project portfolios the authors found a large number of outliers in their sample. However, and here differs our analytical approach, Kulk et al. (2009) excluded almost half their sample because of outlyingness in order to run their logistic regression analysis. In our analysis, however, we are particularly interested in these data points.

## The statistical difference between mediocristan and extremistan

Classical, Gaussian statistics developed and first rose to fame in demographics; a flavour of statistics employed to manage populations by enumerating them (Hacking 1991; Foucault 2009). Demographics love the normal distribution - the idea that the average type is characterised by a measure of centrality, while conformity is measured in standard deviations (Hart 2010). As Hart (2010) further argues the key assumption behind Gaussian statistics is randomness - taking a snapshot of a random sample. Even time series do nothing more than stitching a series of snapshots together. Random sampling philosophically and practically implies that everyone has an equal chance of ending up in the sample - a truly democratic, egalitarian, and atomistic premise (Hart 2010).

Recently the attention of statisticians shifted from Gaussian statistics to Power laws. Power laws were initially discovered in geography describing city sizes (Auerbach 1913 and Zipf 1949) and first used in an economic setting to study the of wealth concentration (Pareto 1964). Power laws have been established for a wide array of natural phenomena. Newman (2005) lists among others word frequency, citations of academic papers, visits to a website, or the magnitude of earthquakes. Power laws were introduced into organisational studies to explain the size distribution of business firms (Simon & Bonini 1958).

Power Laws have two major implications for the understanding of risk in project management (1) exponential proportionality and (2) scale invariance.

Firstly, power laws postulate that the probability of an event happening is exponentially proportional to its impact. Table 1 shows the odds of encountering a millionaire when living in Europe a distribution that follows a quadratic power law. That means doubling the net worth reduces the odds to a quarter. To put it in other words — a millionaire that is twice as rich will occur four times more rarely (Hart 2010).

Table 4 Power Law Example

| **Net worth** | **Probability** |
|---|---|
| >1 million | 1/62.6 |
| >2 million | 1/250 |
| >4 million | 1/1,000 |

Secondly, Power Laws are scale invariant. The relative probability to observe an event of a given size and an event ten times larger is independent of the reference scale (Bouchaud 2001). The scale invariance implies that the thickness of the tales does not change with the impact of the event observed, in our sample the correlation between risk and cost overruns is only r = 0.093 with p < 0.001.

Power Laws have the beauty of relying only on the singular parameter α.

$$P(X \geq x) = \left(\frac{x_0}{x}\right)^{\alpha} = Cx^{-\alpha}$$

The constant α is called the exponent of the power law, whereas C is only introduced to the formula so that the CDF sums up 1 (Newman 2005). The higher moment m is defined as

$$m^{th} = \frac{\alpha x_0^{m-\alpha}}{s^{\alpha+1}}$$

thus moments beyond the $\alpha^{th}$ do not exist. With typical values of alpha being between 1 and 2 only the first moment μ is finite; second order moments and beyond (standard deviation, skewness, kurtosis) do not converge for power law distributions (Axtell 2006).

Taleb (2007) argues that Power Laws are the best way to mathematically describe Black Swans. Black Swans have two key characteristics (a) they are high impact rare events, and (b) they can only be identified retrospectively not prospectively. High impact, rare events are the outcome of thick tails where outliers do not disappear behind a Gaussian average but show a very strong influence on outcomes. The ex-ante unpredictability follows from the fact that the typical Black Swan event will always be unknown, which is statistically tied to the non-convergence of the second, third, and fourth moment. Taleb paints the picture of two worlds: Extremistan and Mediocristan. Extremistan is the world of wild randomness, the world of Power Laws. Mediocristan, on the other hand, is the world of mild randomness (Taleb 2007). Mild randomness allows Gaussian statistics to work because in this world outliers disappear in the averages of large samples (Mandelbrot & Taleb 2006).

*Testing Newtonian Laws vs. Power Laws*

As suggested by (Newman 2005) we calculated the cumulative distribution function (CDF) instead of binning the frequencies to preserve as much distributional information as possible. We calculated the CDF in a way that P(x) equals the probability of a cost overruns equal or greater to x, i.e. the so called survival function

$$P(x) = \int_x^{+\infty} p(x')dx'$$

Figure 1 shows the log-log plot of the CDF. Power laws express as straight linear functions in log-log plots. The chart indicates a drastic change in the rate of change below and above actual/forecast = 100 (value of 2 on the log10 scale).

**Figure 1 Log-Log Chart of the DCF**

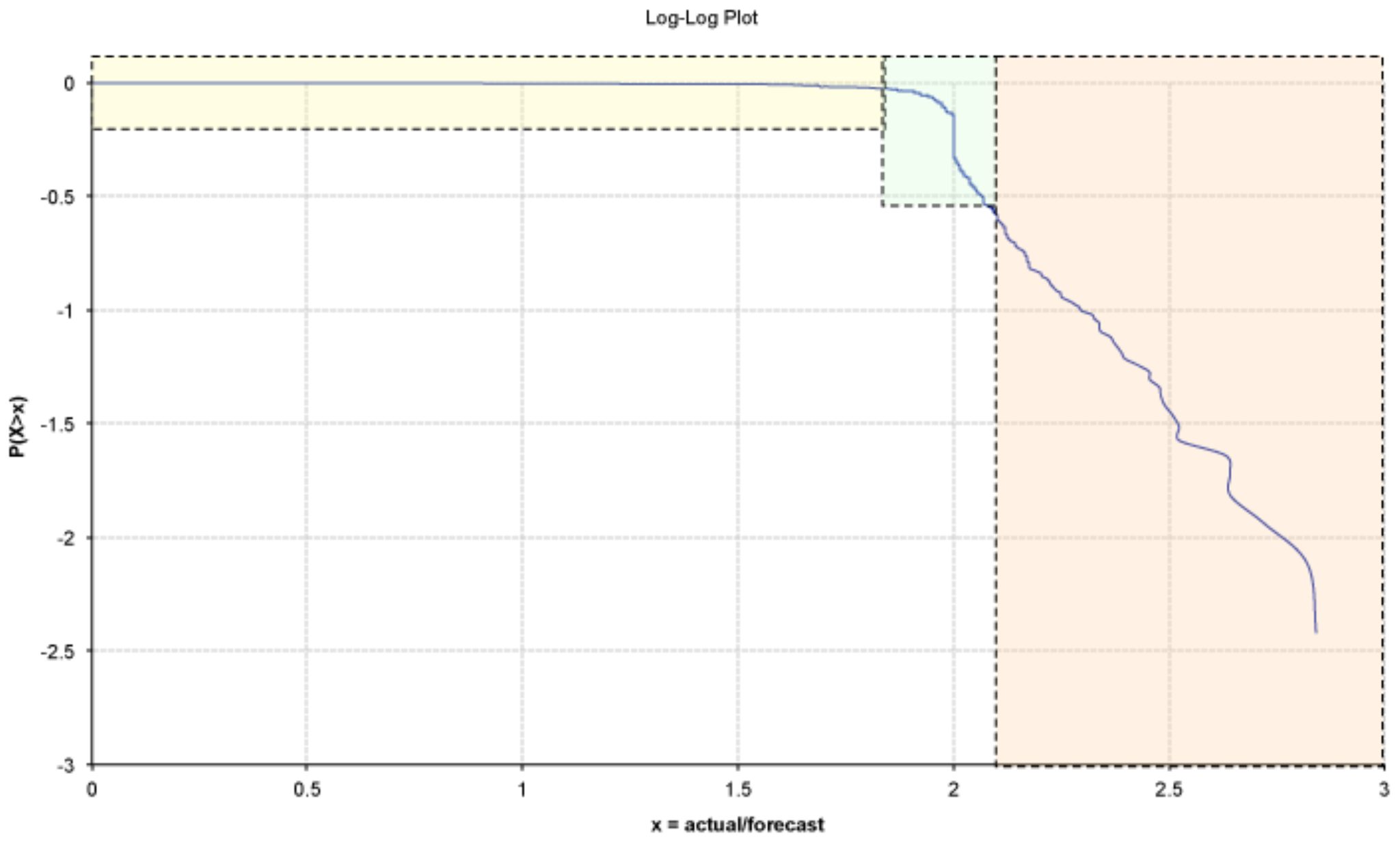

The curve estimation needs to establish four parameters – (1) the slope of the left tail, (2) the mean of the performance regime, (3) the standard deviation of the performance regime, and (4) the slope of the right tail. Furthermore the bounds of the three regimes need to be estimated.

While an iterative EM algorithm is the best method to establish these parameters. This working paper used a first manual approximation to shows the three bounded regimes and estimate the four parameters. Table 5 gives the initial non-iterated results for the sample.

Table 5 The Three Rgimes

| | Political Regime | Performance Regime | Black Swan Regime |
|---|---|---|---|
| Lower Bound | -100% | -30% | +48% |
| Upper Bound | -30% | +48% | +∞ |
| Probability | 6% | 77% | 17% |
| Slope | -0.001 | | -2.1 |
| Mean | | 103.58 | |
| Std. dev. | | 14.79 | |
| F | 160.278 | | 2274.264 |
| p (Fit) | <0.01 | | <0.01 |
| R squared | 0.92 | | 0.98 |

Table 5 shows the fitting results for the three regimes (1) political intent follows a power law, (2) managed performance follows a Gaussian distribution, and (3) Black Swans follow again a power law.

*Political Intent in the Left Tail*

Truth is not in numbers, as Hatherly, Leung & MacKenzie (2008) argued based on the observation of similar kinks in the distribution of reported profits in annual reports. The authors showed that accountants do not passively report the numbers they find but that accounting actively influences the numbers, which leaves visible traces in frequency distributions.

In this analysis the left tail follows a power law with a very low rate of change. A clear indication that this performance outcome is the result of a non-stochastic process. The fit is highly significant and fits 92% of the variance in the log-log plot. The probability of falling prey to budget cuts is relatively low but affects 6% of all projects.

*Managed performance around being on budget*

The managed performance regime shows that normal projects have an average cost overrun of +3.6%, with a standard deviation of 14.8%. This regime has the highest probability of occurring a project stays within normal bounds of (-30%, +48%) with a 77% likelihood.

*Black Swans in the right tail*

The right tail follows a very strong power law the slope is estimated to be -2.1 thus $\alpha = 2.1$. The existence of power laws in the right tail is an expression of an underlying probabilistic generator. The probability of becoming a Black Swan is estimated at 17% - a very high risk compared to thin-tailed distributions where outliers are happening with no more than 0.7% probability on both ends of the tail.

Black Swans are characterised as high impact events with wild randomness (Mandelbrot & Taleb 2006). The power law and the fat tails express the wild randomness. Table 3 shows the impact of Black Swans.

Table 6 Impact of Black Swans

| Risk | All projects | With cost overrun | Black Swans |
|---|---|---|---|
| Cost Risk | +27% | +70% | +197% |
| Schedule Risk | +55% | +60% | +68% |

Firstly, the grand average of the total sample, which shows the average risk of project is 27% in terms of budget, 55% in terms of schedule.

Secondly, following the argument proposed by Yetton (personal conversation, May 12th, 2010) that ICT projects cannot fully be managed like a financial investment portfolio because they are of strategic importance and usually not easily abandoned or divested from (for an in-depth discussion of escalation of commitment see Keil et al., 2000 or Drummond, 1999). Thus a portfolio of ICT projects is more impacted by downside risks of cost overruns and schedule slippage and there is only limited ability to profit from pooling risks and offsetting overruns with budget under-runs and earlier than expected implementations. The sub-sample of the projects with just the downside cost risk of > 0% contains 47% of projects of the total sample; the average cost overrun for this sample almost triples to 70%, schedule slippage increases slightly to 60%.

Thirdly, for the Black Swans the average cost risks almost triples again to +197%, schedule risk slightly increases further to +68%. It is noteworthy that even though an average can be calculated from the empirical data the expectation value for the exponential distribution does not converge. Thus the expected Black Swan project does not exist statistically.

The growing literature on power laws discusses five generating concepts (Mitzenmacher 2004) that might advance the understanding of excessive cost overruns and risk management practices in ICT projects – (1) Preferential Attachment, (2) Optimisation, (3) Proportionate Effects, (4) Monkeys typing randomly, and (5) Double Pareto distributions.

Firstly, preferential attachment describes the web graph, that is the link structure between web pages. The more popular a note is in terms of out-bound connections the more popular it tends to be in terms of in-bound connection. Similarly, (Simon 1955) pointed out the larger a city the greater its ability to attract more population. Preferential attachment also explains book sales, website visits, and species distribution (Mitzenmacher 2004). In the context of ICT projects preferential attachment can be explained by the escalation of commitment (Keil, Mann & Rai 2000), i.e. the unwillingness to pull the plug of failing projects. If good money is thrown after bad money the preferential attachment leads to cost overruns that resemble power laws.

Secondly, information theoretic optimisation leads to power laws (Mandelbrot 1952). Word frequency follows a power law if the langue is optimised for the average amount of information per unit transmission costs. Optimisation also explains power laws behind computer file sizes, the internet hardware infrastructure, and forest sizes (Mitzenmacher 2004). ICT projects are per definition optimised investments. ICT projects aim to deliver an average amount of benefits at the lowest total project costs. Thus we should be able to observe that budget and benefit sizes, as well as cost and benefit risks distributions follow power laws.

Thirdly, multiplicative processes create proportionate effects. Proportionate effects are also known as Gibrat's law. Gibrat's law was the first formalisation of the winner-takes in market entry it has been describes as: "the probability that the next opportunity is taken up by any particular active firm is proportional to the current size of the firm" (John Sutton 1997, p.43). Proportionate effects have also been shown in income distributions, growth of sites on the web, and file sizes (Mitzenmacher 2004). In the context of ICT projects the concept of scope creep can be conceptualised by proportionate effects; the larger the scope of a project, the higher the number of change requests and thus scope creep. Similarly, resource scarcity in organisations creates ICT project portfolios that are highly skewed and follow power law distributions, very few megaprojects are surrounded by a plethora of small projects.

Fourthly, in response to optimisation generators, random walks have been proposed as an alternative explanation. Word frequency distributions can solely be achieved by a random walks, i.e. monkeys typing randomly (Mitzenmacher 2004). In terms of ICT project this explanation is the most dystopian and shows that perhaps none of the capabilities proposed in many project management methodologies actually influences the outcome of the project, similar to the argument by Denrell (Jerker Denrell 2004, J. Denrell 2005).

Fifthly, the combination of two random variables, so called double Pareto distributions, creates power laws in the upper tail (Mitzenmacher 2004). This class of generators has been proposed to account for age effects, i.e., describe the income distribution of a person while accounting for the person's age (ibid.). For ICT projects (Barseghyan 2009) has shown that combining two random thin-tail processes for (1) the productivity of a developer and (2) the difficulty of the development task leads to a fat-tailed development duration distribution.

## Conclusion — Double whammy indeed!

This paper finds that ICT project performance is fooled by complex, random stochastic processes, on one hand, and screwed by social and political constructed number games on the other hand. As Taleb (2001) sarcastically argued managers are fooled by randomness; i.e. a spotless track records of delivering one successful ICT projects after another should be attributed to the ludic fallacy or Black Swan Blindness instead of the superior abilities of the manager. Moreover, delivering a project significantly under budget is hardly an achievement, because the data indicates that in most cases the budget was taken away from the project.

Our first statistical approximation of the collected sample has shown that on average ICT projects perform reasonably well — +27% cost overrun, +55% schedule overrun in three out of four projects. Apart from the risk of getting the budget cut a very high risk exists that a project turns into a Black Swan. One in six projects (17%) with cost overruns of nearly +200% and schedule slippage of nearly 70%.

However, the high over-incidence of Black Swans underlines that ICT projects are a very important source of uncertainty in an organisation. The owner of a portfolio of ICT change initiatives needs to critically assess where the organisation stand when one in six projects develop into a Black Swan with 200% or more cost overruns and schedule delays of 70%.

Lastly, one immediate application for the power law behaviour of cost overruns of ICT projects is the forecasting of failure. When it comes to managing the costs of an ICT project Earned Value Management is currently considered best practice (Jeffrey & Leliveld 2004) and has become a widely accepted standard (Project Management Institute 2008). Earned value methodology originally proposed a linear forecast of costs. That is actual/plan performance is scaled linearly over the progress (Quentin Fleming 2005) later different methods (e.g., Monte Carlo simulations) have been proposed to generate S-curved or logistic forecasting models (e.g., Byung-Cheol Kim & Kenneth F. Reinschmidt 2010). However the discovery of power laws in the right tail of the cost risk distribution indicates that a more drastic approach is needed: if a project encounters a cost overrun of 20% at half time, the most reasonable estimate is not just twice the overrun, but rather the overrun to the power of 2.1 so not to expect 40% but 540% overrun at the end.

## Limitations and future research

While the paper offers a novel contribution to analyse outliers as being the phenomena and not the noise, the findings have limitations. Firstly, since we relied on organisations that measured ICT project performance in terms of cost and schedule, and to a lesser extend also benefits; the sample might be biased towards a certain proficiency level of overall project management. Secondly, the proxy used to test the complexity is less then perfect and might need refinement. Project management scholars are currently developing perception-based scales to measure complexity (e.g. Whitty & Maylor 2009). Thirdly, the geographic focus might introduce cultural biases in the sample selection that prevent generalisability to Asian organisations.

The findings show the importance of rare, high impact events when it comes to managing ICT projects and associated risks for organisations, portfolios, and investments. The findings are limited to a very first discovery of the existence of Black Swans in this particular reference class of risk and uncertainty.

The paper points towards many open questions for future research. Among those are - What is the relation of Black Swans to project complexity? Which determinants change the probability of Black Swan Events? Which organisational processes can be linked to the theoretical underlying generators, e.g., preferential attachment?

Remarks: This is a working paper and thus represents work in progress, we are currently further enlarging the sample size to 1,800 projects. We also work on improving the curve fitting and regime estimation by using an EM algorithm.

## References

Anon, 1999. Chaos, Standish Group Inc.

Auerbach, F. 1913. Das Gesetz der Bevölkerungskonzentration. Petermanns Geographische Mitteilungen 59, 74–76.

Axtell, R. L. 2001. Zipf Distribution of U.S. Firm Sizes. Science 293(September), 1818–1820.

Axtell, R. L. 2006. Firm Sizes: Facts, Formulae, Fables And Fantasies. CSED Working Paper No. 44, February 2006. Washington, D.C.

Barseghyan, P. 2009. Human Effort Dynamics and Schedule Risk Analysis. PM World Today 11(3).

Boehm, B., 1991. Software Risk Management: Principles and Practices. IEEE Software, 8(1), pp.32–41.

Bouchaud, Jean P., 2001. Power laws in economics and finance: some ideas from physics. Quantitative Finance 1, pp. 105-112.

Bowker, G. & Star, S.L., 1994. Knowledge and infrastructure in international information management: Problems of classification and coding. In L. Bud-Frierman, ed. Information acumen: The understanding and use of knowledge in modern business. London: Routledge, pp. 187–213.

Brooks, F.P., 1975. The mythical man-month: Essays on software engineering, Reading, Mass: Addison-Wesley.

Bruijn, H. de & Leijten, M., 2008. Management characteristics of mega-projects. In H. Priemus, B. Flyvbjerg, & B. van Wee, eds. Decision-making on mega-projects: Cost-benefit analysis, planning, and innovation. Cheltenham, UK: Elgar, pp. 23–39.

Brys, G., Hubert, M. & Struyf, A., 2004. A robust measure of skewness. Journal of Computational and Graphical Statistics, 13(4), pp.996–1017.

Budescu, D.V. & Bruderman, M., 1995. The relationship between the illusion of control and the desirability bias. Journal of Behavioural Decision Making, 8(2), pp.109–125.

Byung-Cheol K. & Reinschmidt K. F. 2010. Probabilistic Forecasting of Project Duration Using Kalman Filter and the Earned Value Method. Journal of Construction Engineering and Management 136(8), 834.

Campbell, D.T. & Fiske, D.W., 1959. Convergent and discriminant validity by the multitrait-multimethod matrix. Psychological Bulletin, 56(2), pp.81–105.

Chell, E., 2004. Critical Incident Technique. In C. Cassell & G. Symnon, eds. Essential guide to qualitative methods in organizational research. London: Sage Publ., pp. 45–60.

Conte, S.D., Dunsmore, H.E. & Shen, Y.E., 1986. Software engineering metrics and models, Redwood City: Benjamin-Cummings Publishing.

Crewe, E. & Young, J., 2002. Bridging Research and Policy: Context, Evidence and Links, London, UK.

DeMarco, T., 1995. Why does software cost so much? And other puzzles of the Information Age, New York, NY: Dorset House Publ.

DeMarco, T. & Lister, T., 1999. Peopleware: Productive projects and teams 2nd ed. New York, NY: Dorset House Publ.

Demski, J. S. & Feltham, G. A. .1978. Economic Incentives in Budgetary Control Systems. The Accounting Review, 53(2), pp. 336-359.

Dempster, A.P., Laird, N.M. & Rubin, D.B., 1977. Maximum Likelihood from Incomplete Data via the EM Algorithm. Journal of the Royal Statistical Society, 39, pp.1–38.

Denrell, J., 2004. Random Walks and Sustained Competitive Advantage. Management Science 50(7), pp. 922-934.

Denrell, J., 2005. Should We Be Impressed With High Performance? Journal of Management Inquiry 14(3), pp. 292-298.

Dermer, J.D. & Lucas, R.G., 1986. The illusion of managerial control. Accounting, Organizations and Society, 11(6), pp.471–482.

Dospisil, J. & Polgar, T., 1994. Conceptual modelling in the hypermedia development process. In Proceedings of the 1994 ACM SIGCPR conference. Proceedings of the 1994 ACM SIGCPR conference. New York: ACM Press, pp. 97–104.

Douglas, M. & Wildavsky, A., 2010. Risk and culture: An essay on the selection of technological and environmental dangers 1st ed. Berkeley, Calif.: Univ. of California Press.

Drummond, H., 1999. Are We Any Closer to the End? Escalation and the Case of Taurus. International Journal of Project Management, 17(1), pp.11–16.

Dutton, J.E. & Webster, J., 1988. Patterns of Interest around Issues: The Role of Uncertainty and Feasibility. The Academy of Management Journal, 31(3), pp.663–675.

Elahi, S., 2011. Here be dragons … exploring the “unknown unknowns.” Futures, 43(2), pp.196–201.

Erev, I., Wallsten, T. S. & Budescu, D. V., 1994. Simultaneous over-and under-confidence: The role of error in judgment processes. Psychological Review 101(3), 519–527.

Eveleens, L.J. & Verhoef, C., 2009. Quantifying IT forecast quality. Science of Computer Programming, 74(11-12), pp.934–988.

Eveleens, L.J. & Verhoef, C., 2010. The Rise and Fall of the Chaos Report Figures. IEEE Software, 27(1), pp.30–36.

Fleming, Q., 2005. Earned value project management. 3rd. Newtown Square PA: Project Management Institute.

Flyvbjerg, B., 2005. Beyond the planning fallacy: reference class forecasting in practice (draft paper in progress).

Flyvbjerg, B., 2007. Eliminating Bias In Early Project Development through Reference Class Forecasting and Good Governance. In K. J. Sunnev r a g, ed. Beslutninger pa svakt informasjonsgrunnlag. Trondheim, Norway, pp. 90–110.

Flyvbjerg, B., 2006. From Nobel Prize to Project Management: Getting Risks Right. Project Management Journal, 37(3), p.5.

Flyvbjerg, B., Holm, M.K.S. & Buhl, S.L., 2005. How (in)accurate are demand forecasts in public works projects? The case of transportation. Journal of the American Planning Association, 71(2), pp.131–146.

Flyvbjerg, B., Bruzelius, N. & Rothengatter, W., 2006. Megaprojects and risk: An anatomy of ambition 4th ed. Cambridge: Cambridge Univ. Press.

Flyvbjerg, B., Holm, M.K.S. & Buhl, S.L., 2002. Underestimating costs in public works projects: Error or lie? Journal of the American Planning Association, 68(3), pp.279–295.

Fortune, J. & White, D., 2006. Framing of Project Critical Success Factors by a Systems Model. International Journal of Project Management, 24, pp.53–65.

Foucault, M., 2009. Governmentality, in Graham Burchell, u.a. (Hg.): The Foucault effect. Chicago, Ill: Univ. of Chicago Press.

Friedman, T. & van Decker, J.E., 2010. 2009 Gartner FEI Technology Study, Florham Park, N.J.: Financial Executive International & Gartner Inc.

Gaddis, P.O., 1959. The Project Manager. Harvard Business Review, 37(3), pp.89–98.

Glass, R., 2006. The Standish report: does it really describe a software crisis? Communications of the ACM, 49(8), pp.15–16.

Hacking, I., 1991. The Taming of Chance. Cambridge: Cambridge University Press.

Haimes, Y.Y., 2009. Risk modeling, assessment, and management 3rd ed. Hoboken, NJ: Wiley.

Hart, K., 2010. Models of statistical distribution: A window on social history. Anthropological Theory 10(1-2), pp. 67-74.

Hatherly, D., Leung, D. & MacKenzie, D., 2008 The Finitist Accountant: Classifications, Rules and the Construction of Profits, in Living in a Material World: Economic Sociology meets Science and Technology Studies, edited by Trevor Pinch and Richard Swedberg (Cambridge, MA: MIT Press, 2008), 131-160. Refereed.

Heaton, J.B., 2002. Managerial optimism and corporate finance. Financial Management, 31(2), pp.33–45.

Hobday, M., 2000. The project-based organisation: An ideal form for managing complex products and systems? Research Policy, 29(7-8), pp.871–893.

Huchzermeier, A. & Loch, C.H., 2001. Project Management Under Risk: Using the Real Options Approach to Evaluate Flexibility in R&D. Management Science, 47(1), pp.85–101.

Janis, I.L., 1971. Groupthink. Psychology Today, 1971, pp. 43-46.

Jeffrey, M. & Leliveld, I., 2004. Best Practices in IT Portfolio Management. MIT Sloan Management Review 45(3), pp. 41–49.

Jick, T.D., 1979. Mixing Qualitative and Quantitative Methods: Triangulation in Action. Administrative Science Quarterly, 24(4), pp.602–611.

Jones, C., 2008. Applied Software Measurement: Global Analysis of Productivity and Quality 3rd ed. McGraw-Hill.

Jones, C., 2003. Software assessments, benchmarks, and best practices 2nd ed. Boston, Mass.: Addison-Wesley.

Jones, T.C., 1998. Estimating software costs, New York: McGraw-Hill.

Jorgensen, M., 2004. A review of studies on expert estimation of software development effort. Journal of Systems and Software, 70(1-2), pp.37–60.

Jorgensen, M. & Molokken-Ostvold, K., 2006. How large are software cost overruns? A review of the 1994 CHAOS report. Information and Software Technology, 48(4), pp.297–301.

Kahneman, D. & Lovallo, D., 1993. Timid Choices and Bold Forecasts: A Cognitive Perspective on Risk Taking. Management Science, 39(1), pp.17–31.

Kahneman, D. & Tversky, A., 1977. Intuitive Prediction: Biases and Corrective Procedures, Eugene, Oregon.

Keil, M., 1995. Pulling the Plug: Software Project Management and the Problem of Project Escalation. MIS Quarterly, 19(4), pp.421–447.

Keil, M. & Flatto, J., 1999. Information systems project escalation: a reinterpretation based on options theory. Accounting, management, and information technologies, 9(2), pp.115–139.

Keil, M. & Mann, J., 1997. Understanding the Nature and Extent of IS Project Escalation: Results from a Survey of IS Audit and Control Professionals, Los Alamitos, Calif.: IEEE Computer Soc. Press.

Keil, M., Tan, B.C.Y., Wei, K.-K., Saarinen, T., Tuunainen, V. & Wassenaar, A., 2000a. A Cross-Cultural Study on Escalation of Commitment Behavior in Software Projects. MIS Quarterly, 24(2), pp.299–325.

Keil, M., Mann, J. & Rai, A., 2000b. Why Software Projects Escalate: An Empirical Analysis and Test of Four Theoretical Models. MIS Quarterly, 24(4), pp.631–664.

Klein, C.T.F. & Helweg-Larsen, M., 2002. Perceived control and the optimistic bias: A meta-analytic review. Psychology and Health, 17(4), pp.437–446.

Kulk, E., 2009. IT Risks in Measure and Number, PhD thesis.

Kulk, G.P., Peters R.J & Verhoef, C., 2009. Quantifying IT estimation risks. Science of Computer Programming, 74(11-12), pp.900–933.

Langer, E.J., 1975. The Illusion of Control. Journal of Personality and Social Psychology, 32(2), pp.311–328.

Laughhunn, D.J., Payne, J.W. & Crum, B., 1980. Managerial Risk Preferences for Below-Target Returns. Management Science, 26(12), pp.1238–1249.

Leach, L.P., 2007. Critical Chain Project Management. In P. W. G. Morris & J. K. Pinto, eds. The Wiley guide to managing projects. Hoboken, NJ: John Wiley & Sons, p. Chapter 33.

Lecky, P., 1945. Self-Consistency: A Theory of Personality, Washington, DC: Island Press.

Linehan, C. & Kavanagh, D., 2006. From Project Ontologies to Communities of Virtue. In S. Cicmil & D. Hodgson, eds. Making Projects Critical. Basingstoke, England: Palgrave Macmillan, pp. 51–67.

Mandelbrot, B. 1952. An Informational Theory of the Statistical Structure of Languages, in Jackson, Willis (Hg.): Communication theory: Papers read at a symposium on applications of communication theory, held at the Institution of Electrical Engineers, London, September 22nd-26th, 1952. Woburn, MA: Butterworths Scientific Publications, 486–502.

Mandelbrot, B. & Taleb, N. N., 2006. A focus on the exceptions that prove the rule. URL: http://viewpointsofacommoditytrader.com/wp-content/uploads/2009/08/A-focus-on-the-exceptions-that-prove-the-rule.pdf

March, J.G. & Shapira, Z., 1987. Managerial Perspectives on Risk and Risk Taking. Management Science, 33(11), pp.1404–1418.

McFarlan, F.W., 1981. Portfolio approach to information systems: Assessing the risk of their projects, separately and in the aggregate, will help managers make more informed decisions and ensure more successful outcomes. Harvard Business Review, (September-October), pp.142–150.

Meszaros, J.R., 1999. Preventive Choices: Organizations' Heuristics, Decision Processes and Catastrophic Risks. Journal of Management Studies, 36(7), pp.977–998.

Mitzenmacher, M., 2004. A Brief History of Generative Models for Power Law and Lognormal Distributions. Internet Mathematics 1(2), pp. 226–251.

Molokken-Osvold, K. & Jorgensen, M. eds. 2003. A review of software surveys on software effort estimation // Proceedings: 30 September - 1 October 2003, Villa Mondragone, University of Rome "Tor Vergata“,” Roman Castles (Rome), Italy. In K. Molokken-Osvold & M. Jorgensen, eds. A review of software surveys on software effort estimation // Proceedings: 30 September - 1 October 2003, Villa Mondragone, University of Rome "Tor Vergata“,” Roman Castles (Rome), Italy. A review of software surveys on software effort estimation // Proceedings: 30 September - 1 October 2003, Villa Mondragone, University of Rome "Tor Vergata“,” Roman Castles (Rome), Italy. Los Alamitos, Calif: IEEE Computer Society.

Montealegre, R. & Keil, M., 2000. De-Escalating Information Technology Projects: Lessons from the Denver International Airport. MIS Quarterly, 24(3), pp.417–447.

Mumford, E., 1995. Effective systems design and requirements analysis: The ETHICS approach, Basingstoke: Macmillan.

National Audit Office, 2008. Department for Work and Pensions: Information Technology Programmes National Audit Office, ed. London.

Newman, M., 2005. Power laws, Pareto distributions and Zipf‘s law. Contemporary Physics 46(5), pp. 323-351.

Packendorff, J., 1995. Inquiring into the Temporary Organization: New Directions for Project Management Research. Scandinavian Journal of Management, 11(4), pp.319–333.

Powell, R., 2008. Measuring Extreme Financial Risk with Power Laws. Bank Accounting & Finance, p.p. 35-40.

Project Management Institute (PMI), 2008. A guide to the project management body of knowledge: (PMBOK guide). 4. ed. Newton Square, Pa.: Project Management Inst.

Sauer, C. & Cuthbertson, C., 2004. The State of IT Project Management in the UK 2002-2003. Computer Weekly Website. Available at: http://www.computerweekly.com/Articles/2010/10/12/243318/CW+-The-state-of-IT-project-management-in-the-UK.htm.

Sauer, C., Gemino, A. & Reich, B.H., 2007. The Impact of Size and Volatility on IT-Project Performance: Studying the Factors Influencing Project Risk. Communications of the ACM, 50(11), pp.79–84.

Simon, H. A. 1955., On a class of skew distribution functions. Biometrika 42(3-4), 425–440.

Simon, H. A. & Bonini, C. P., 1958. The Size Distribution of Business Firms. The American Economic Review 48(4), 607–617.

Stahl, B.C., Lichtenstein, Y. & Mangan, A., 2006. The Limits of Risk Management: A Social Construction Approach. In B. C. Stahl, ed. Communications of the International Information Management Association (IIMA). Las Vegas, pp. 15–22.

Stahlknecht, P. & Hasenkamp, U., 2005. Einführung in die Wirtschaftsinformatik Elfte, vollständig überarbeitete Auflage., Berlin, Heidelberg: Springer Berlin Heidelberg.

Staw, B.M., 1981. The escalation of commitment to a course of action. The Academy of Management Review, 6(4), pp.577–587.

Stirling, A., 2010. Addressing Uncertainty, Ambiguity and Ignorance in Sustainability Appraisal, Presentation to interdisciplinary workshop on “Cost Benefit Analysis: Uncertainty, Discounting and the Sustainable Future”, Eindhoven Technical University, 12-13th April, 2010.

Sutton, J., 1997. Gibrat‘s Legacy. Journal of Economic Literature 35(1), pp. 40-59.

Taleb, N. N., 2001. Fooled by randomness: The hidden role of chance in the markets and in life. New York: Texere.

Taleb, N.N., 2005. The Roots of Unfairness: The Black Swan in Arts and Literature. Literary Research/Recherche Litteraire, 21(41-42), pp.241–254.

Taleb, N.N., 2007. The black swan: The impact of the highly improbable, New York, NY: Random House.

Taleb, N.N., 2007. Black Swans and the Domains of Statistics. The American Statistician 61(3), pp. 198-200.

Taleb, N.N. Common Errors in Interpreting the Ideas of the Black Swan and Associated Papers. URL: http://ssrn.com/abstract=1490769.

Verhoef, C., 2002. Quantitative IT~Portfolio Management. Science of Computer Programming, 45(1), pp.1–96.

Webb, E.J. et al., 1966. Unobtrusive measures: Nonreactive research in the social sciences, Chicago: Rand McNally.

Whitty, S.J. & Maylor, H., 2009. And then came Complex Project Management (revised). International Journal of Project Management, 27(3), pp.304–310.

Wilkinson, A. & Ramirez, R., 2010. Canaries in the Mind: Exploring How the Financial Crisis Impacts 21st Century Future-Mindfulness. Journal of Futures Studies, 14(3), pp.45–60.

Williams, T.M., 1992. Practical Use of Distributions in Network Analysis. The Journal of the Operational Research Society, 43(3), pp.265–270.

Zhang, G.P., Keil, M., Rai, A. & Mann, J., 2003. Predicting information technology project escalation: A neural network approach. European Journal of Operational Research, 146, pp.115–129.

Zipf, G.K, 1949. Human Behaviour and the Principle of Least Effort. Reading, Mass: Addison-Wesley.

Zmud, R.W., 1980. Management of Large Software Development Effort. MIS Quarterly, 4(2), pp.45–55.

{The Standish Group International, Inc.}, 2001. Extreme Chaos.